\documentclass[runningheads]{llncs}

\usepackage[T1]{fontenc}
\usepackage{graphicx}

\usepackage{cite}

\usepackage{adjustbox}
\usepackage{makecell}
\usepackage[utf8]{inputenc}

\usepackage{bbding}

\usepackage{svg}

\usepackage{csquotes}

\usepackage{color}
\usepackage{hyperref}

\usepackage{orcidlink}
\renewcommand{\orcidID}[1]{\,\orcidlink{#1}}

\newcommand\samethanks[1][\value{footnote}]{\footnotemark[#1]}

\begin{document}

\title{Market-Analysis-Driven Methodology for Assesing Charging Station Cybersecurity}
\titlerunning{Charging Station Cybersecurity Assessment Methodology}

\author{Jakob L\"{o}w\thanks{Both authors contributed equally to this work.}\orcidID{0009-0006-7088-8684} \and
Lukas Eder\samethanks \orcidID{0009-0008-1579-3855} \and
Alexander M\"{u}ller \and
Hans-Joachim Hof\orcidID{0000-0002-6930-9271}}

\authorrunning{J. L\"{o}w et al.}

\institute{Technische Hochschule Ingolstadt, Security in Mobility,\\
Ingolstadt, Germany\\
\email{\{jakob.loew,lukas.eder,hof\}@thi.de}}


\maketitle

\begin{abstract}
Modern charging communication standards for electric vehicles include optional security controls such as TLS-based authentication and encryption.
However, with tens of thousands of fast charging points deployed in any given country, individually testing each one for security control support is infeasible.
This paper proposes a scalable, extrapolation-based methodology for assessing charging station cybersecurity at a national level.
A market analysis identifies operator-manufacturer pairs, enabling the targeted selection of charging stations for field testing, whose results can then be extrapolated to all stations sharing the same combination.
We demonstrate this methodology for Germany, covering over 40000 CCS charging points as of December 2025.
With a manageable number of field tests, our extrapolated data examines 51.9\% of german CCS charging stations. It shows that only 27.4\% of charging stations in our scope provide TLS-protected communication, despite widespread theoretical support.

\end{abstract}

\section{Introduction}
Electric vehicles are becoming increasingly popular in Germany and the rest of the world \cite{noauthor_openevcharts_2025}.
While many users charge their vehicles at home or at work, fast charging stations are required on long-distance trips in order to minimize break times.
Fast charging stations rely on high-level communication to keep the high-power charging process within safe limits, preventing battery damage and degradation.
While this high-level communication enables convenience features such as automatic digital payment handled by the vehicle and the charging station, it also results in a larger attack surface for charging communication.
Recent papers have demonstrated the feasibility of man-in-the-middle attacks \cite{dudek_v2g_2019, eder_charging_2025, szakaly_dception_2026} as well as payment fraud \cite{conti_evexchange_2022, loew_pnc_2026} against charging communication.
Although current charging communication standards already include security controls that provide session authentication and could prevent such attacks, their use is currently optional.
At present, there is no overview of what percentage of charging stations in the field today actually support these optional security controls.

Given the scale and heterogeneity of modern charging infrastructures, exhaustive security testing of all deployed charging stations is impractical. To address this challenge, we introduce an extrapolation-based methodology that leverages market structure and targeted measurements to assess the cybersecurity posture of charging infrastructure at national scale.
By performing a market analysis, specific charging stations are selected, allowing extrapolation from a small number of tested stations to an overview of the current cybersecurity posture of charging communication.
The methodology is based on two assumptions, which are validated later.
Most charging point operators do not manufacture their own charging hardware.
While there are thousands of fast charging stations in the field today, operated by hundreds of operators, the vast majority were built by a small number of manufacturers. The first assumption is therefore that charging stations manufactured by the same company support the same set of features, e.g. support for transport layer security (TLS).
The second assumption is that two charging stations built by the same manufacturer and operated by the same company use the same configurations and thus support the same set of security controls.
Together, these assumptions imply that testing one charging station allows the test results to be extrapolated to all stations built by the same manufacturer and operated by the same company.

We validate these assumptions and demonstrate the methodology for Germany, where it enables a qualified assessment of security control availability for more than half of all charging stations using a manageable number of field tests.

Our analysis reveals a significant gap between the security mechanisms defined in standards and their deployment in practice. While most charging station hardware supports TLS-based secure communication, only a minority of stations in the field actually enable and successfully use these mechanisms. This indicates that the current security posture of charging infrastructure is primarily limited by operational and configuration challenges rather than technical capabilities.

\section{Charging Communication Security} 
In order to charge an electric vehicle, AC power from the grid has to be converted to an adequate DC voltage for charging the vehicle battery.
In addition, safety parameters such as maximum current, temperature constraints, and voltage limits need to be followed.
For fast charging stations, this AC to DC conversion happens outside of the vehicle, inside the charging station.
Therefore, the vehicle needs to be able to communicate with the charging station and exchange safety parameters.

\begin{figure}[ht]
    \centering
    \includesvg[width=0.99\linewidth]{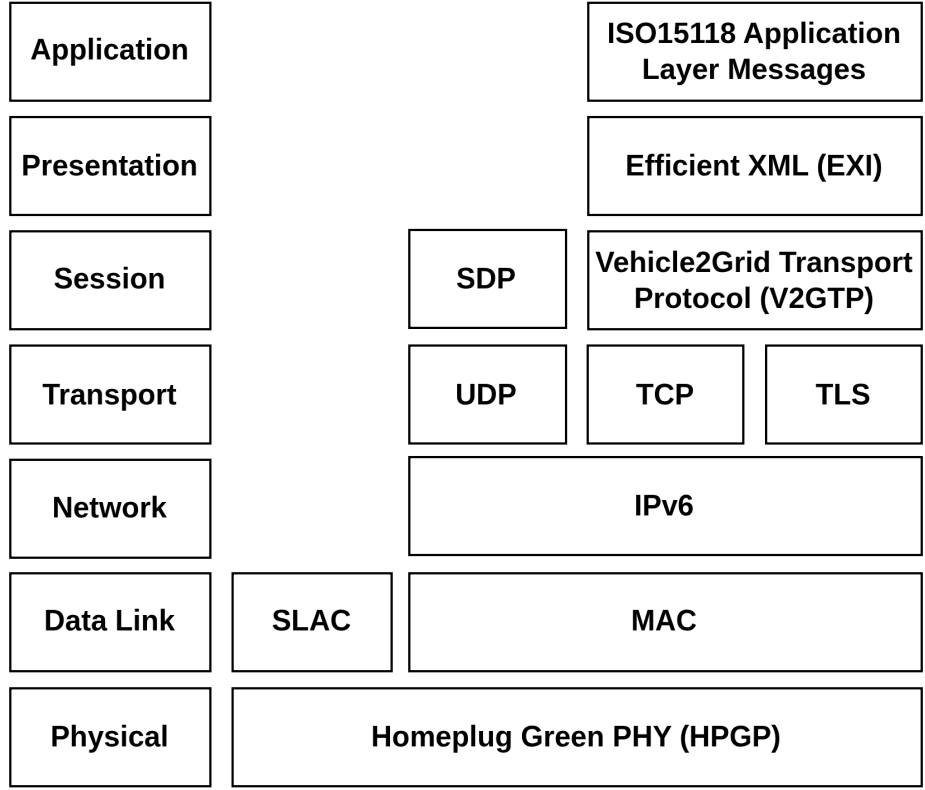}
    \caption{Charging Communication OSI Layer Overview}
    \label{fig:iso15118-osi}
\end{figure}

The leading standard for charging communication in Europe is ISO 15118 \cite{isoiec_isoiec_2012, isoiec_isoiec_2012-1, isoiec_isoiec_2012_20}.
The standard builds on common communication protocols such as powerline communication, IPv6, and Transport Layer Security.
Figure \ref{fig:iso15118-osi} shows the communication technology used in ISO 15118 for each of the open systems interconnect (OSI) model layers.
The initialization steps, namely signal level attenuation characterization (SLAC) and the service discovery protocol (SDP), are designed specifically for charging communication.
SLAC is used to form a joint powerline network between the vehicle and the charging station.
Afterwards, SDP is used to exchange network addresses and to decide on the optional use of TLS for encrypting the main charging communication.
Since SLAC messages are transmitted in plain text and SDP messages do not include means of message authentication, the initialization steps of ISO 15118 open up the possibility of misuse scenarios \cite{dudek_v2g_2019, dudek-homeplugav-2015, low_fast_2024, loew_pnc_2026, szakaly_dception_2026}.
When both the vehicle and the charging station signal support for TLS, the main charging communication can be authenticated and encrypted using state-of-the-art algorithms.
When TLS is used, the ISO 15118 standard also enables a special certificate-based automatic payment scheme.
The use of TLS prevents potential attackers from impersonating a communication party or eavesdropping on the communication, significantly improving the security posture of a charging communication session.

\begin{figure}[ht]
    \centering
    \includesvg[width=0.99\linewidth]{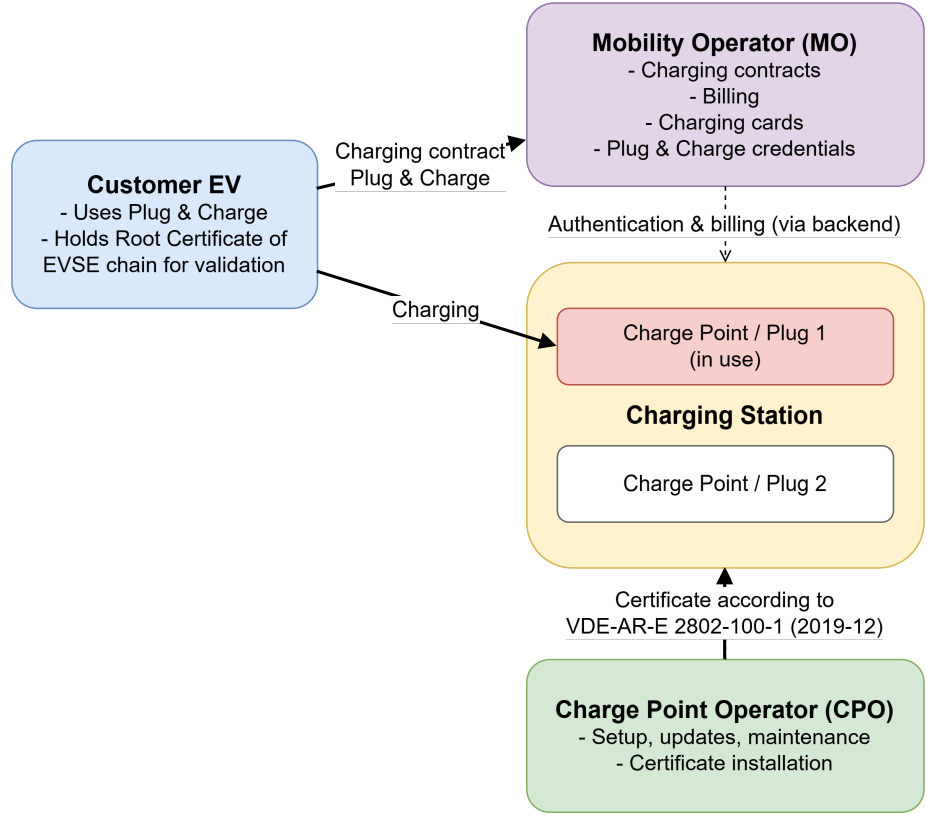}
    \caption{Simplified overview of EV charging participants}
    \label{fig:ev-charging-overview}
\end{figure}

For clear terminology throughout the paper, Figure \ref{fig:ev-charging-overview} shows the participants in the charging process relevant to this work. Charging stations can have one or more charging points, i.e., can serve one or more vehicles at the same time.
For comparability, we count individual charging points rather than charging stations, since stations can have varying numbers of points depending on manufacturer, model, and hardware version.
Charging stations are often part of charging parks, each run by one charge point operator (CPO).
Operators run multiple parks, resulting in larger operators operating hundreds or even thousands of charge points. The charging stations operated by a CPO also have a manufacturer, which is generally not the CPO itself. Most CPOs do not manufacture charging stations themselves; instead, they buy off-the-shelf models. Another participant in the charging process is the mobility operator (MO), which handles everything related to billing. Multiple roles (MO, CPO, manufacturer) can be held by the same party or company, but this is not required.

\section{Related Work}
Prior surveys and threat analyses have examined the attack surface of ISO 15118 and CCS-based fast charging \cite{lee_study_2014, bao_threat_2018, low_fast_2024}.
At the powerline and physical layer, concrete attacks have been demonstrated, including practical exploitation of HomePlug AV \cite{dudek-homeplugav-2015}, powerline-based command injection against V2G charging \cite{dudek_v2g_2019}, passive PHY-layer eavesdropping of CCS sessions \cite{baker_losing_2019}, wireless denial-of-service via signal interference \cite{kohler_brokenwire_2023}, and wireless SDR-based man-in-the-middle attacks including TLS stripping \cite{szakaly_dception_2026}.
A separate line of work targets the powerline modems themselves: these have been shown to expose configuration and data-extraction interfaces over the powerline \cite{szakaly_short_2025}, a primitive that has subsequently been leveraged to install filter and firewall rules enabling sniffing and man-in-the-middle attacks on the charging session \cite{eder_charging_2025}.
At the application layer, relay attacks against the charging protocol \cite{conti_evexchange_2022} and plug-and-charge payment-authentication relays \cite{loew_pnc_2026} have been used to commit payment fraud.
While most of these works propose new countermeasures alongside the vulnerabilities they identify, the basic security mechanisms already specified in the charging standards -- most importantly TLS for authentication and encryption -- are themselves only optional and require support from both the vehicle and the charging station to take effect.
Whether deployed charging stations actually enable these standardized mechanisms is an empirical question that has so far received only limited attention; this paper argues that closing the deployment gap for these existing controls is at least as pressing as proposing new ones.

Szakály et al. \cite{szakaly_current_2025} present the closest prior measurement study.
The authors tested almost 400 charging stations across the United Kingdom, Switzerland, and Hungary for supported communication protocols and security controls.
Their sample is opportunistic rather than stratified: stations were measured as accessible, without targeted selection by charge point operator or hardware manufacturer.
As a result, their work establishes that TLS deployment varies significantly between operators and documents concrete configuration shortcomings, but it cannot quantify how prevalent each configuration is within a given country or region.
The methodology proposed in this paper is designed for exactly this question: by stratifying field tests along the (CPO, manufacturer) dimensions that, as shown in Section~\ref{sec:assumption-validation}, govern configuration in practice, we extrapolate from a small sample to national-scale deployment estimates.
The two approaches are complementary; we in fact incorporate 25 of the charging stations measured by Szakály et al.\ into our assumption-validation set in Section~\ref{sec:assumption-validation}, where their cross-country measurements of overlapping CPOs strengthen the case that configurations are consistent within a (CPO, manufacturer) pair across borders.


\section{Market Analysis Methodology}
The proposed methodology combines a market analysis of the target region with targeted field testing and extrapolation to assess the cybersecurity state of charging infrastructure at scale.
The goal is to identify all charging stations, their manufacturers, and their operators, in order to select a minimal set of stations for field testing that maximizes extrapolation coverage.
We demonstrate this methodology for Germany using openly available data sources, as described in the following subsections.

Extrapolating from a limited number of field-tested charging stations to a larger population requires two assumptions to hold:

\begin{itemize}
    \item A1: Charging Stations of the same manufacturer support the same set of features, while the CPO controls which features are enabled through configuration and certificate provisioning.
    \item A2: CPOs use the same configuration on all Charging Stations from the same manufacturer.
\end{itemize}

The validity of these assumptions is empirically evaluated in Section \ref{sec:assumption-validation} based on measurements across multiple manufacturers, models, installation years, and CPOs.


\subsection{Extrapolation-Based Assessment Methodology}

The methodology consists of the following steps:

\noindent
Step 1 - Data Acquisition:
Publicly available datasets are collected to obtain charging point locations, charge point operators (CPOs), and charging station manufacturers and models. In this work, data from the Bundesnetzagentur \cite{bundesnetzagentur_anzahl_2024} and goingelectric.de \cite{noauthor_dokumentation_nodate} are used.\\

\noindent
Step 2 - Data Preprocessing:
The collected data is cleaned and normalized. This includes unifying inconsistent manufacturer labels (e.g., \textit{Tesla} vs.\ \textit{Supercharger}) and removing incomplete entries. The resulting dataset contains only charging points with known CPO and manufacturer assignments.\\

\noindent
Step 3 - Clustering:
Charging points are grouped into clusters based on their CPO and manufacturer:
\[
\texttt{Cluster} = (\texttt{CPO}, \texttt{Manufacturer})
\]
Each cluster is assumed to represent a homogeneous configuration domain.\\

\noindent
Step 4 - Sampling Strategy:
For each cluster, a subset of charging stations is selected for field testing. The selection aims to maximize coverage of charging points while minimizing the number of required tests. Priority is given to clusters with a large number of charging points.\\

\noindent
Step 5 - Field Testing:
Selected charging stations are analyzed using the proposed EV-side testing device. The following properties are evaluated:
\begin{itemize}
    \item availability of TLS,
    \item supported communication protocols (DIN SPEC 70121, ISO 15118-2),
    \item presence of valid certificate chains.
\end{itemize}

\noindent
Step 6 - Extrapolation:
The results obtained from tested charging stations are extrapolated to all charging points within the same cluster. This is based on the assumption that charging stations within a cluster share identical capabilities and configuration.\\

\noindent
Step 7 - Aggregation:
The extrapolated results are aggregated across all clusters to estimate the overall security posture of the charging infrastructure, including TLS deployment rates and protocol distribution at national scale.\\

\noindent
This structured approach enables scalable assessment of large charging infrastructures while requiring only a limited number of field measurements.


\subsection{Data Collection} 
The proposed extrapolation of charging station tests requires quantified knowledge about charging points, manufacturers, and charge point operators.
While official sources such as the charging station list provided by the federal network agency \cite{bundesnetzagentur_ladekarte_nodate} exist, these include only operator and connector information and lack details about the hardware used.
Therefore, the official data must be enriched with additional sources in order to obtain a fully featured dataset.

The German website goingelectric.de hosts a community of electric vehicle enthusiasts, including a forum, a wiki, and a curated map of charging stations.
In addition to charging locations, users can add pictures and metadata to each charging point, such as supported charging power, manufacturer, and model.
While the dataset includes information about some charging points in other countries, its main focus is on charging stations in Germany.
As part of this research, we used the official goingelectric API as well as web scraping to download information such as charging station manufacturers and models.

Since the information about charging station manufacturers and models is added manually by individuals, the data required some manual sanitization.
For example, some charging stations built and operated by Tesla are labeled as \textit{Tesla}, while others are labeled as \textit{Supercharger}.
Out of 114078 CCS charging points present on goingelectric.de in December 2025, 44313 (or 39\%) are located in Germany.
Of these 44313 charging points, another 3364 (or 7.6\%) do not have a charging station manufacturer labeled, leaving 40949 charging points for analysis.

For each of these charging points, the charging station manufacturer, the model, the CPO operating the station, and the corresponding mobility operator were extracted.
In most cases, the charge point operator is identical to the mobility operator handling the billing of charging sessions.
However, some mobility operators do not operate all charging stations themselves, but instead allow third parties to register charging stations within their network. %
The largest of these is \textit{NewMotion}, which used to allow private individuals to register their publicly accessible charging stations, enabling others to use their infrastructure in exchange for a share of the profit.
While in theory this means that charging stations in the \textit{NewMotion} network could differ in configuration, this mainly applies to slow AC charging stations.
DC fast charging stations included in the \textit{NewMotion} network are mainly operated by Shell, which acquired \textit{NewMotion} in 2021.
Additionally, there are at least three other mobility operators that do not run charging stations themselves, but allow third parties to register stations in their network: \textit{be.energised}, \textit{Ladenetz}, and \textit{innogy}.
None of these networks advertise support for plug and charge, which makes the availability of security controls such as TLS in those charging stations unlikely.
We later validate this assumption with some of the CPOs that use these networks.


\label{sec:data-summary}
\subsection{Data Summary} 
The collected data shows a large number of small and medium-sized mobility and charge point operators.
Many traditional companies active in the energy sector, such as \textit{EnBW}, \textit{EWE}, \textit{allego}, \textit{innogy}, \textit{Vattenfall}, \textit{E.ON}, and \textit{TEAG}, have invested in the electric vehicle charging market.
Automotive players are also among the top charge point operators, including \textit{Tesla}, \textit{Volkswagen} (as \textit{Elli}), and various other carmakers through \textit{IONITY}.
Retail chains such as \textit{Lidl}, \textit{Aldi}, \textit{Kaufland}, and \textit{Edeka} have built charging stations at their existing locations.
Lastly, players from the fossil energy sector, such as \textit{Aral} with \textit{Aral Pulse} and \textit{Shell} through \textit{NewMotion}, are also among the top mobility operators in Germany.

\begin{figure}[t]
    \centering
    \includegraphics[width=0.7\linewidth]{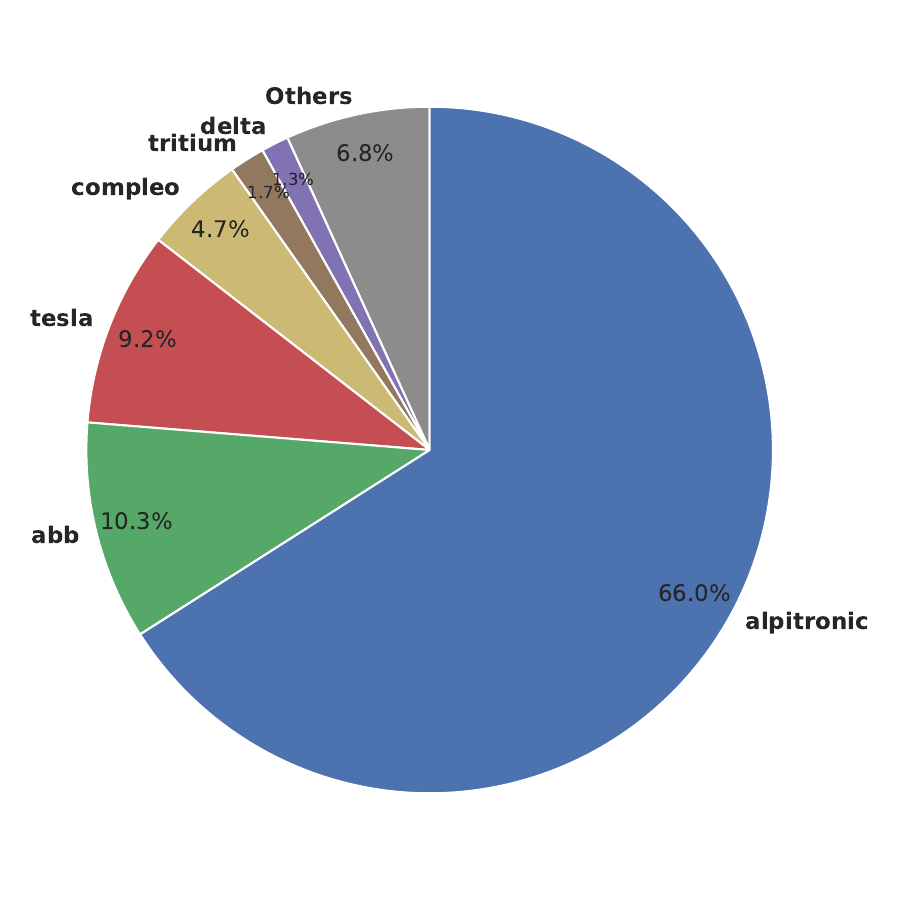}
    \caption{Charging Station Manufacturers Market Share}
    \label{fig:manufacturers}
\end{figure}

Regarding charging station manufacturers, the collected data shows that nearly 66\% of CCS charging points belong to charging stations built by a single manufacturer: \textit{Alpitronic}.
Both \textit{ABB} and \textit{Tesla} hold around a 10\% market share of installed charge points in Germany.
The remaining 14.5\% are split among multiple smaller manufacturers.
Figure \ref{fig:manufacturers} shows an overview of the manufacturer market share among all CCS charging points in Germany.

Many charge point operators use \textit{Alpitronic} charging stations almost exclusively.
As shown in Table \ref{fig:network_manufacturers}, four of the five leading CPOs use Alpitronic charging stations for more than 90\% of their charging points; the fifth is Tesla, which builds its charging stations itself.
This concentration simplifies the planned security analysis of charging stations in Germany.

\label{sec:assumption-validation}
\subsection{Validation of Extrapolation Assumptions}

The proposed methodology relies on two assumptions: (A1) charging stations from the same manufacturer provide the same set of communication capabilities, and (A2) charging stations operated by the same CPO are configured consistently within each manufacturer group. The data used for the validation can be found in the repository linked at the end of section \ref{sec:test-results}.

\paragraph{Validation of A1 (manufacturer capability consistency).}
To evaluate (A1), charging stations from multiple manufacturers (Alpitronic, ABB, Tritium, Compleo, Efacec, EKOenergetyka, ChargePoint) were analyzed across different model generations and installation years (2018--2024). In all observed cases, charging stations from the same manufacturer supported the same set of communication capabilities at the protocol level.

In particular, no case was observed in which two charging stations from the same manufacturer differed in their capability to support TLS. However, differences in TLS activation were observed and are attributed to operator-side configuration and certificate provisioning.

Additionally, analysis of publicly available firmware documentation \cite{alpitronic_gmbh_hyperdoc_nodate} indicates that different models of a major manufacturer share a common software platform and configuration interface, further supporting this assumption.

\paragraph{Validation of A2 (CPO configuration consistency).}
To evaluate (A2), multiple charging stations from the same CPO--manufacturer combinations were tested across different locations and installation years. In total, 16 charging stations from three major CPOs were analyzed:
\begin{itemize}
    \item EnBW (Alpitronic): 8 stations across 5 locations
    \item Aral (Alpitronic, Compleo): 5 stations across 4 locations
    \item IONITY (Alpitronic, ABB, Tritium): 3 stations across 3 locations
\end{itemize}

Across all tested stations within each CPO--manufacturer pair, identical behavior regarding TLS support and supported protocols was observed, independent of installation year (2019--2025) and hardware model. This suggests a high degree of configuration standardization within each CPO.

To further support this observation, measurement data from \cite{szakaly_current_2025} was incorporated. For overlapping CPOs (e.g., IONITY, Allego, Fastned), consistent configurations were observed across different European countries. This adds 25 additional charging stations (ABB and EKOenergetyka, 2018--2022) to the validation set.

\paragraph{Limitations.}
The validation is based on a finite number of sampled charging stations and may not capture all sources of variation. In particular, the following factors may influence the validity of the assumptions:
\begin{itemize}
    \item configuration changes over time (e.g., firmware updates or certificate rollouts),
    \item regional differences within large CPOs,
    \item temporary misconfigurations or partial deployments.
\end{itemize}

Therefore, the extrapolation results should be interpreted as an approximation of the current deployment state rather than an exact measurement. Nevertheless, the high consistency observed across all tested samples suggests that large-scale charging networks are operated with highly standardized configurations, supporting the feasibility of the proposed methodology.

\begin{table}[ht]
    \caption{Charging Station Manufacturers per Network}
    \label{fig:network_manufacturers}
    \centering
    \begin{tabular}{l c c l c c}
        CPOs & Points & Points (\%) & Manufacturer & Points (\%) \\
        \hline
        enbw & 7593 & 18.5 & alpitronic  & 96.1 \\
        & & & abb & 3.0 \\
        tesla & 3770 & 9.2 & tesla  & 100.0 \\
        aral & 2887 & 7.1 & alpitronic  & 92.8 \\
        & & & volkswagen & 6.9 \\
        ewe & 1999 & 4.9 & alpitronic  & 99.3 \\
        allego & 1860 & 4.5 & alpitronic  & 95.3 \\
        & & & efacec & 4.2 \\
        pfalzwerke & 1662 & 4.1 & alpitronic  & 82.9 \\
        & & & enercharge & 8.1 \\
        & & & abb & 6.5 \\
        & & & siemens & 2.2 \\
        newmotion & 1573 & 3.8 & alpitronic  & 59.8 \\
        & & & abb & 35.0 \\
        & & & volkswagen & 5.0 \\
        be.energised & 1454 & 3.6 & alpitronic  & 67.6 \\
        & & & ads-tec & 13.3 \\
        & & & compleo & 6.5 \\
        & & & abb & 6.1 \\
        & & & ekoenergetyka & 1.1 \\
        ionity & 1363 & 3.3 & tritium  & 48.3 \\
        & & & abb & 32.9 \\
        & & & alpitronic & 18.8 \\
        ladenetz & 1355 & 3.3 & alpitronic  & 58.2 \\
        & & & compleo & 15.9 \\
        & & & abb & 8.7 \\
        & & & enercharge & 7.7 \\
        & & & ads-tec & 4.0 \\
        & & & siemens & 2.0 \\
        & & & delta & 1.0 \\
    \end{tabular}
\end{table}


\section{Charging Station Testing Device} 
To analyze the supported communication protocols and security mechanisms of CCS charging stations under real-world conditions, a dedicated charging station testing device was developed. The device is designed to emulate the EV-side communication behavior required for standardized high-level charging communication, while deliberately excluding any energy transfer functionality. This enables systematic, reproducible, and safe measurements of protocol support, TLS availability, and certificate deployment across different electric vehicle supply equipments (EVSE) without interacting with high-voltage components. The following subsections describe the hardware and software architecture of the testing device in detail.

\subsection{Hardware Setup}
We designed our testing device to be portable. It emulates the EV side required for charging communication analysis. The system can be operated in a fully mobile manner, requiring only the test device itself and an external power bank, which enables flexible deployment independent of fixed laboratory infrastructure.

A Raspberry Pi Zero 2 W serves as the central processing unit of the test device. It executes the complete communication logic and is responsible for recording and logging all exchanged data. To enable HomePlug Green PHY (HPGP) power line communication, a Codico RED-BEET-EVAL-BOARD-E 2.0 power line modem is integrated into the system. The modem is electrically connected to the Raspberry Pi via an existing SPI interface and a power connection, providing a compact and efficient communication link between both components.

The physical interface to the EVSE is realized using a modified Type 2 charging cable. Since the test device is exclusively intended for communication analysis, no DC power contacts are required. The communication process is intentionally terminated before the EVSE would enable any power transfer, thereby ensuring that no voltage is applied during operation.

Within the Type 2 connector, a set of resistors combined with switches is implemented to accurately reproduce the different CP-states as specified by the relevant standards. Furthermore, a diode is integrated into the connector to suppress negative voltage components of the PWM signal on the control pilot line. The PP resistor is statically implemented in the connector to represent a fixed cable current rating. The CP and PE conductors are directly connected to the power line communication modem.

Overall, the proposed hardware setup provides a complete emulation of the EV-side communication components required for charging while omitting high-power and high-voltage elements. This approach enables a portable analysis of the communication behavior between EV and EVSE in a controlled manner.


\subsection{Software Setup}
The charging station testing device implements a vehicle-side simulation of standardized charging communication between an EV and an EVSE for testing purposes. The simulated communication supports DIN SPEC 70121, ISO 15118-2, and ISO 15118-20, thereby enabling a systematic and reproducible assessment of protocol compliance and security-related properties of charging stations.

The software strictly follows the normative communication sequence defined by the standards. This sequence comprises the SLAC coupling procedure, the service discovery protocol, the optional establishment of a secured communication channel using TLS, and the subsequent supported application protocol negotiation. These phases collectively determine the protocol stack, security mechanisms, and application profiles that are supported and selected for a charging session.

To systematically evaluate EVSE behavior under different communication and security configurations, the test process is executed in multiple independent test cycles. In each cycle, the simulated EV advertises a predefined set of communication capabilities and security requirements during the service discovery and protocol negotiation phases. The tested scenarios include:
\begin{enumerate}
    \item Service discovery requests that require TLS exclusively, allowing the detection of TLS support and the availability of certificates at EVSE-side 
    \item Protocol negotiation requests for ISO 15118-2 without the use of TLS
    \item Protocol negotiation requests without TLS that advertise all supported standards, including DIN SPEC 70121, ISO 15118-2, and ISO 15118-20
    \item Protocol negotiation requests without TLS that advertise only DIN SPEC 70121
\end{enumerate}

The first test allows the detection of TLS support and the availability of certificates on the EVSE side. Tests 1, 2, and 4 check for HLC communication protocol availability. Test 3 checks for the EVSE's preferred communication protocol.

All information relevant for protocol selection and security evaluation -- including the supported and selected communication standards, the availability of secured communication, and the use of digital certificates -- is exchanged during these early stages of the communication process. As no additional information required for the intended analysis is negotiated beyond this point, the test process is deliberately terminated after the successful completion of these steps. This controlled termination ensures that the charging session does not proceed to energy transfer, resulting in the desired behavior that the EVSE does not output charging voltage.

\label{sec:test-results}
\section{Test Results}
Table \ref{fig:results} summarizes the observed protocol and security feature support across the analyzed CCS charging stations.
As expected for CCS EVSEs, all tested charging stations implement DIN SPEC 70121 while most also support ISO 15118-2. In contrast, the practical use of TLS is limited.
Although ISO 15118-2 allows TLS for secure communication, only a small fraction of the measured charging stations successfully established TLS-protected sessions. In most cases, TLS was either disabled or not operational.
Our extrapolation covers 51.9\% of German charging stations and specifically shows that only 27.4\% of them provide TLS-secured charging communication.

The measurement results suggest that the lack of TLS deployment is not caused by technical limitations of the charging hardware, but by challenges in certificate provisioning and trust management.
This conclusion is supported by the observation that charging stations from the same product families -- and in several cases even newer hardware revisions -- exhibited differing TLS behavior.
Specifically, EVSE models that did not establish TLS were observed alongside equivalent or older models from the same product line that successfully supported TLS, indicating that hardware capabilities alone do not explain the absence of TLS in practice.

For TLS to be accepted by the vehicle, the EVSE certificate chain must terminate in a root CA trusted by the EV.
While on the web, obtaining certificates for encrypted communication is free and easy to automate, in the charging domain PKI providers usually charge a one-time fee as well as recurring fees for each charging station.
This effectively hinders the use of TLS for secure charging communication.

At the time of writing, at least four operators of V2G public key infrastructures are known to provide trusted root CAs for ISO 15118 \cite{charin_charin_nodate, hubject_download_nodate, nexusgroup_identities_nodate, irdeto_irdeto_nodate}.
However, within our dataset, only certificates issued under the Hubject V2G Root CA \cite{hubject_download_nodate} were observed.
All other stations either operated without TLS or failed to present a certificate.

TLS in ISO 15118 is mandatory when using plug and charge.
Therefore, all networks that support plug and charge payment also support TLS.
IONITY has supported plug and charge since October 2021 and appears to have updated chargers built before that date, providing them with the certificates required for successful TLS communication.
Similarly, Aral introduced plug and charge in early 2022.
Charging stations built before this introduction also support TLS, indicating updates after initial construction.

In general, support for TLS seems to be mainly driven by the introduction of plug and charge payment rather than security concerns.
While in our dataset 27.4\% of charging spots support TLS, we assume the real-world number to be even lower.
Due to the required effort and upfront cost, plug and charge is currently mainly supported by large charging point operators.
While our dataset includes both large and small operators, the distribution is not proportional to the total number of charging points installed.
Small CPOs left out of the dataset are expected to have an even lower rate of support for TLS.

Apart from the extracted results shown in table \ref{fig:results} we are also publishing all raw data gathered from various charging stations including communication dumps as well as model information on github:
\url{https://anonymous.4open.science/r/charging-station-tests-germany-28EC/}

\renewcommand{\arraystretch}{1.7}
\begin{table}
\centering
\begin{tabular}{ l c | l c | c | l c c c }
CPO & CPO \% & OEM & OEM \% & Cluster \% & Model & Year & \makecell{ISO \\ 15118-2} & TLS \\ 
\hline
EnBW & 18.5 & Alpitronic & 95.9 & 17.7 & HYC400 & 2023 & \CheckmarkBold & \XSolidBold \\
 &  & & & & HYC300 & 2019, 2025 & \CheckmarkBold & \XSolidBold \\
 &  & & & & HYC200 & 2024 & \CheckmarkBold & \XSolidBold \\
 &  & & & & HYC150 & 2021, 2023 & \CheckmarkBold & \XSolidBold \\
Aral & 7.1 & Alpitronic & 92.8 & 6.6 & HYC300 & 2021, 2023 & \CheckmarkBold & \CheckmarkBold \\
 &  & Compleo & 0.1 & 0.0 & \makecell{Cito \\ BM 500} & 2022 & \XSolidBold & \XSolidBold \\
EWE & 4.9 & Alpitronic & 99.3 & 4.9 & HYC300 & 2021, 2022 & \CheckmarkBold & \XSolidBold \\
allego & 4.5 & Alpitronic & 95.2 & 4.3 & HYC300 & 2022 & \CheckmarkBold & \CheckmarkBold \\
pfalzwerke & 4.1 & Alpitronic & 82.9 & 3.4 & HYC150 & 2021 & \CheckmarkBold & \XSolidBold \\
IONITY & 3.3 & Tritium & 48.3 & 1.6 & Veefil PK & 2019 & \CheckmarkBold & \CheckmarkBold \\
 & & ABB & 32.9 & 1.1 & HP CP500 & 2024 & \CheckmarkBold & \CheckmarkBold \\
 & & Alpitronic & 18.8 & 0.6 & HYC400 & 2023 & \CheckmarkBold & \CheckmarkBold \\
Lidl & 2.4 & ABB & 79.2 & 1.9 & Terra 60 & 2021 & \CheckmarkBold & \XSolidBold \\
Elli & 2.3 & Compleo & 95.7 & 2.2 & \makecell{Cito \\ BM 500} & 2022 & \CheckmarkBold & \XSolidBold \\
Mer & 2.0 & Alpitronic & 90.8 & 1.8 & HYC150 & 2021 & \CheckmarkBold & \XSolidBold \\
Aldi & 2.0 & Alpitronic & 98.8 & 2.0 & HYC150 & 2021 & \CheckmarkBold & \XSolidBold \\
Kaufland & 1.7 & ABB & 92.4 & 1.6 & Terra 54 & 2022 & \CheckmarkBold & \XSolidBold \\
Edeka & 1.1 & Compleo & 84.1 & 0.9 & \makecell{Cito \\ BM 500} & ? & \XSolidBold & \XSolidBold \\
fastned & 0.8 & Alpitronic & 100 & 0.8 & HYC300 & 2021 & \CheckmarkBold & \XSolidBold \\
Circle K     & 0.7 & ABB & 65.0 & 0.5 & HP CP500 & 2021 & \XSolidBold & \XSolidBold \\
\hline
\textbf{\% Of All} & 55.4 & & & 51.9 & & & 48.7 & 14.2 \\
\makecell{\textbf{\% Of } \\ \textbf{Clusters}} & & & & 100 & & & 93.8 & 27.4 \\
\end{tabular}
\caption{Test Results for various CPOs}
\label{fig:results}
\end{table}


\section{Conclusion}
Our research shows a significant lack of support for security controls in chargers deployed in the field today.
While current research continues to identify vulnerabilities and shortcomings in charging communication and to propose new security controls to prevent attacks, in reality most chargers do not support even basic authentication or encryption.
While some operators have made the effort to support TLS, this appears to be a prerequisite for introducing plug and charge rather than a proactive step to improve the cybersecurity of the charging infrastructure.
The current state of cybersecurity in charging communication appears to resemble that of the web years ago.
Making TLS mandatory for all charging sessions could improve the confidentiality and integrity of charging communication.

The proposed market-analysis-driven methodology is not limited to Germany and can be applied to any country or region where charging station data is available, enabling comparable assessments of charging infrastructure cybersecurity elsewhere.

Future research could evaluate current certificate distribution techniques and investigate how a certificate enrollment process similar to the Automatic Certificate Management Environment (ACME) used for web servers could be integrated into the charging ecosystem.
Similar to the service provided by Let's Encrypt, an easy, automatic, and free way to obtain certificates could significantly accelerate the adoption of authentication and encryption in charging communication.
Additionally, such an approach could help bring secure-by-default practices to charging communication, rather than requiring network operators to manually obtain and configure certificates for each charging station individually.

\begin{credits}
\subsubsection{\ackname}
\textit{redacted for anonymized version}
\end{credits}

\bibliographystyle{splncs04}
\bibliography{references}

\end{document}